\documentclass[%
 superscriptaddress,
 amsmath,amssymb,
 aps,
 prb,
 twocolumn,
 longbibliography 
]{revtex4-2}

\usepackage{graphicx}
\usepackage{dcolumn}
\usepackage{bm}
\usepackage{xcolor}
\usepackage{tikz}
\usepackage{xfrac}
\usepackage{blindtext}
\usepackage{times}
\usepackage{multirow}
\usepackage{chemformula}
\usepackage{braket}

\usepackage{color}
\usepackage[colorlinks=true, urlcolor=blue, linkcolor=blue, citecolor=blue, pdftex]{hyperref}

\usepackage[normalem]{ulem}

\definecolor{darkblue}{HTML}{004D6B}
\definecolor{darkred}{HTML}{8c1515}
\definecolor{darkgreen}{HTML}{006400}
\usepackage{hyperref}
\hypersetup{
	colorlinks=true,
	urlcolor=darkred,
	citecolor=darkblue,
	linkcolor=darkred,
	breaklinks
}

\usepackage[caption=false]{subfig}

\graphicspath{{figures/}}

\newcommand{\be}{\begin{equation}}
\newcommand{\ee}{\end{equation}}

\newcommand{\bea}{\begin{eqnarray}}
\newcommand{\eea}{\end{eqnarray}}
\newcommand{\beal}{\begin{align}}
\newcommand{\eeal}{\end{align}}

\newcommand{\intg}{\mathbb{Z}}
\newcommand{\abs}[1]{\left| #1 \right|} 

\newcommand{\CoNbO}{CoNb$_2$O$_6$} 

\begin{document}

\title{Two-dimensional coherent spectroscopy of CoNb$_{\mathbf{2}}$O$_{\mathbf{6}}$} 

\author{Yoshito Watanabe}
\affiliation{Institute for Theoretical Physics, University of Cologne, 50937 Cologne, Germany}

\author{Simon Trebst}
\affiliation{Institute for Theoretical Physics, University of Cologne, 50937 Cologne, Germany}

\author{Ciar\'an Hickey}
\affiliation{School of Physics, University College Dublin, Belfield, Dublin 4, Ireland}
\affiliation{Centre for Quantum Engineering, Science, and Technology, University College Dublin, Dublin 4, Ireland}

\begin{abstract}
With recent advances in terahertz (THz) sources and detection, two-dimensional coherent spectroscopy (2DCS), 
which allows to probe {\it nonlinear} responses in a two-frequency plane,
now reaches the meV regime relevant for quasiparticle excitations in magnetic materials. 
This opens a promising route to reveal many-body phenomena that evade linear-response probes, including quasiparticle interactions, fractionalized excitations, and modes forbidden by (dipolar) selection rules. To date, applications to quantum magnets are still emerging---most experiments have focused on classical magnets, and a solid demonstration in a {\it quantum} magnet has yet to be established. 
Here we present a theoretical study of 2DCS in \CoNbO, a quasi-one-dimensional Ising magnet that is believed to host fractionalized spinons 
which at low temperatures are confined by weak interchain coupling. 
Our analysis, 
which builds on an effective $S=1/2$ Hamiltonian that has been well constrained by inelastic neutron scattering and THz spectroscopy,
is found to reveal unambiguous 2DCS signatures of spinon deconfinement above the low-temperature ordered phase.
Using a four-spinon approximation, we track these 2DCS signatures by sequentially building a faithful microscopic model for \CoNbO, starting from the exactly solvable one-dimensional transverse-field Ising model (1$d$ TFIM) and successively adding interactions to capture its key low-energy physics. In particular, adding a bond-dependent staggered YZ interaction to the 1$d$-TFIM already reproduces many key spectral features of the full material Hamiltonian. Within this TFIM+YZ model, we find a series of bound states, including a \emph{four-spinon bound state}
that is distinct from the familiar two-spinon bound states. We further find that introducing a confinement potential suppresses sharp \emph{spinon-echo} features in the two-frequency space, which are thought to reflect an underlying continuum of fractionalized excitations. Our results provide concrete predictions and clear targets for future THz 2DCS experiments on \CoNbO\  and related quasi-one-dimensional quantum magnets.
\end{abstract}

\maketitle

\section{Introduction}

Quantum magnets offer fertile ground for exploring collective phenomena such as unconventional collective orderings with multipolar excitations,
the formation of long-range entangled quantum spin liquid states, and the emergence of fractionalized excitations  \cite{lacroix2011, Starykh2015, Savary2017}. 
In forming conceptual understanding (quasi-)one-dimensional materials play a particularly important role as their reduced dimensionality not only naturally allows for fractionalized spinon excitations, their underlying microscopic models are also often simple enough to study in great detail by analytical and numerical methods~\cite{giamarchi2003, Mikeska2004}. On the experimental side, much of the progress in this area has been driven by spectroscopic techniques, especially inelastic neutron scattering~\cite{Coldea2001, Lake2005,DallaPiazza2015, Bera2017, Bai2021} and THz spectroscopy~\cite{Kida2009, Kampfrath2011, Wang2016, Chauhan2020, Legros2021, Wang2024}, which have revealed a wide variety of excitations and demonstrated striking agreement between theory and experiment.

Despite these successes, linear-response spectroscopy also has limitations. In particular, it often struggles to unambiguously resolve fractionalized excitations. In many cases, the expected signal appears as a broad continuum~\cite{Lake2010, Lake2010, Lake2013, Han2012, Benton2012, Punk2014, Knolle2015, Banerjee2016, Little2017, Gaudet2019, Gao2019, Dai2021, Zhang_X_2023}, which can be difficult to distinguish from disorder-broadened magnons or from poorly defined excitations in glassy systems~\cite{Shen2016, Zhu2017, Kimchi2018, Ma2018}. To address these challenges, there has been growing interest in two-dimensional coherent spectroscopy (2DCS)~\cite{Lu2017, Wan2019}, a nonlinear technique originally developed in ultrafast multidimensional spectroscopy of molecular and biological systems (e.g., 2D IR and electronic spectroscopy)~\cite{Cho2008, Hamm_Zanni_2011, Gelzinis2019}. By using THz pulses to probe higher-order response functions, 2DCS can access dynamical information beyond the reach of conventional linear spectroscopy, and in principle enables sharper differentiation between different excitation mechanisms~\cite{Parameswaran2020, Choi2020, Li2021, Fava2021, Negahdari2023, Gao2023, Sim2023, Sim2023_2, Qiang2024, Potts2024, Kaib2025, Zhang2024-2, Fava2025, Zhang2024, Zhang2024-2, Huang2024, Zhang2025}.

In this work, we investigate the 2DCS signatures of the quasi-one-dimensional magnetic material \CoNbO~\cite{Maartense1977, Coldea2010, Lee2010, Kinross2014, Morris2014, Liang2015}. 
We provide concrete predictions for its third-order nonlinear response with a two-fold aim in mind: First, to interpret the 2DCS signatures of \CoNbO\ using a well-defined starting point---the deconfined spinon excitations of the ferromagnetic phase of the one-dimensional transverse-field Ising model (TFIM), though it has always been appreciated that in \CoNbO\ finite interchain couplings confine the spinons at low temperatures. Indeed, its excitation spectrum reflects an intimate proximity to the 1$d$-TFIM. 
Second, as a step toward understanding the full 2DCS spectrum of \CoNbO, we analyze a series of intermediate models that extend the 1$d$-TFIM. 
This analysis builds on recent studies~\cite{Fava2020, Morris2021, Woodland2023, Woodland2023-2, Churchill2024, Gallegos2024, Konieczna2025} which have further elucidated the microscopic description of \CoNbO\ to more accurately capture its low-energy behavior by including additional interactions. Linear-response measurements have shown excellent agreement with these theoretical models. In systematically including these additional exchange terms in our analysis as well, we can not only fully decipher the nonlinear response of \CoNbO, but also provide broader insight into the general features of nonlinear spectroscopy in systems beyond exactly solvable limits.
Taken as a whole, this setting makes \CoNbO\ an ideal platform for exploring the potential of 2DCS in an exemplary quantum magnet.

In this step-by-step analysis, many of the key 2DCS features of \CoNbO\ can already be captured by the 1$d$-TFIM with additional YZ interactions, which admits, within its effective spinon Hamiltonian, another exactly solvable point -- the so-called {\it localized} limit -- for which we identify a series of bound-state signatures that are not accessible in linear response. We further discuss several distinctive features of the full spectrum, including anti-diagonal rephasing signals, which have been of particular interest~\cite{Wan2019, Li2021}, and direct signatures of a four-spinon bound state, which we refer to as a {\it tetraquark} state~\cite{Vovrosh2022}, also observed in the localized limit.

The paper is organized as follows. In Sec.~\ref{sec:model}, we introduce the spin Hamiltonian of \CoNbO, and in Sec.~\ref{sec:methods} we briefly review the 2DCS technique and our computational methods. Section~\ref{sec:construction} builds the 2DCS spectra step by step, starting from the exactly solvable 1$d$-TFIM and successively adding the interaction terms present in \CoNbO. Readers primarily interested in the final material-level results may proceed directly to Sec.~\ref{sec:full_spectrum}, where we present the 2DCS spectra of the full Hamiltonian and trace their evolution along an interpolation from the TFIM+YZ model to the full Hamiltonian. Finally, Sec.~\ref{sec:summary_outlook} summarizes our results and discusses experimental considerations and future directions.

\begin{figure}[b]
	\centering
	\includegraphics[width=1.0\columnwidth]{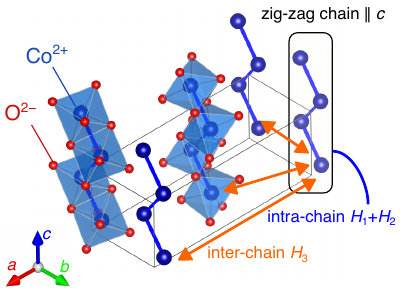}
	\caption{{\bf Quasi-1D Ising compound \CoNbO.} A schematic of the crystal structure is shown, with Nb$^{5+}$ ions omitted for clarity. The magnetic Co$^{2+}$ ions, each surrounded by six O$^{2-}$ ions forming octahedra, build zigzag chains along the $c$ axis through edge sharing. The intra-chain Hamiltonian $\mathcal{H}_1 + \mathcal{H}_2$ is dominated by a ferromagnetic Ising interaction. The effect of the much weaker interchain coupling is represented by the mean-field term $\mathcal{H}_3$, which becomes relevant only below the N\'eel temperature $T_\mathrm{N} \approx 2.9$~K. The structure is drawn using VESTA~\cite{Momma2008, Weitzel1976}.}
	\label{fig:CoNb2O6}
\end{figure}

\section{Model}
\label{sec:model}

\CoNbO\ has been extensively studied as a quasi-one-dimensional Ising magnet, providing access to a rich variety of many-body quantum phenomena. These include the paradigmatic quantum phase transition induced by a transverse magnetic field and the emergence of an $E_8$ spectrum, expected for an integrable {\it massive} field theory, near this one-dimensional quantum critical point, as evidenced by linear-response measurements~\cite{Maartense1977, Coldea2010, Lee2010, Kinross2014, Morris2014, Liang2015}. Indeed, a hierarchy of confined spinon bound states, sometimes called ``mesons'' in analogy to quark confinement, has been observed and is now well established~\cite{Coldea2010, Morris2014, Woodland2023}.

The crystal structure of \CoNbO, as illustrated in Fig.~\ref{fig:CoNb2O6}, is orthorhombic with space group $Pbcn$~\cite{Weitzel1976}. Magnetic Co$^{2+}$ (3$d^7$, $S=3/2$) ions reside in distorted oxygen octahedra that share edges to form zigzag chains along the crystallographic $c$ axis; in the $ab$ plane, the chains form an anisotropic triangular arrangement. The orthorhombic unit cell contains four Co$^{2+}$ ions arranged in two symmetry-related chains. The dominant exchange is an Ising ferromagnetic interaction between nearest neighbors along the chains, while weak interchain couplings produce three-dimensional magnetic order below $T_\mathrm{N}\approx 2.9~\mathrm{K}$~\cite{Scharf1979, Heid1995, Hanawa1994}. Although Co$^{2+}$ carries spin $S=3/2$, spin-orbit coupling and the distorted octahedral crystal field conspire to isolate an Ising-like Kramers doublet, with the low-energy sector well captured by an effective $S=1/2$ pseudospin with an anisotropic $g$-tensor. The full effective $S=1/2$ Hamiltonian can be captured by the sum of three parts

\begin{equation}
\mathcal{H} = \mathcal{H}_1 + \mathcal{H}_2 + \mathcal{H}_3,
\end{equation}
\begin{equation}
\begin{split}
\mathcal{H}_1 &= J \sum_j \left[ - S_j^z S_{j+1}^z - \alpha_\mathrm{S} \left( S_j^x S_{j+1}^x + S_j^y S_{j+1}^y \right)\right., \\
&\left.\quad + (-1)^j \alpha_{yz} \left( S_j^y S_{j+1}^z + S_j^z S_{j+1}^y \right) \right] + \sum_j h_y S_j^y ,\\
\mathcal{H}_2 &= J \sum_j \left[- \alpha_\mathrm{A} \left( S_j^x S_{j+1}^x - S_j^y S_{j+1}^y \right)\right.\\
&\left.\quad + \alpha_{\mathrm{AF}} S_j^z S_{j+2}^z + \alpha_{\mathrm{AF}}^{xy} \left( S_j^x S_{j+2}^x + S_j^y S_{j+2}^y \right)\right],\\
\mathcal{H}_3 &= J \sum_j 2 \alpha_{\mathrm{MF}} \left( \langle S^y \rangle S_j^y - \langle S^z \rangle S_j^z \right) ,\\
\end{split}
\end{equation}
where $h_y = g_y\mu_\mathrm{B}B_y$. In the following, when referring to the strength of the magnetic field, we use the physical field $B_y$ in units of Tesla (T), whereas when discussing excitation energies we use $h_y$ in units of meV.
The local $z$-axis is the Ising direction, tilted by $\pm 31^\circ$ from the crystallographic $c$-axis in the $ac$ plane. The sign of the tilt depends on which chain the site belongs to. The $y$-axis is parallel to the crystallographic $b$-axis, making it a global axis. The local $x$-axis is defined to form the right-handed coordinate system.  

Following the careful determination of coupling parameters in Ref.~\cite{Woodland2023}, we assume the following parametrization throughout $J = 2.48$ meV, $\alpha_\mathrm{S} = 0.251$, $\alpha_{yz} = 0.226$, $g_y = 3.32$, $\alpha_\text{A} = -0.021$, $\alpha_{\mathrm{AF}} = 0.077$, $\alpha_{\mathrm{AF}}^{xy} = 0.031$, and $\alpha_{\mathrm{MF}} = 0.0158$. Such a parametrization is also supported by microscopic derivations of the effective Hamiltonian~\cite{Fava2020,Churchill2024, Gallegos2024,Konieczna2025}.
$\mathcal{H}_1$ thus comprises the most dominant intra-chain interactions---the nearest-neighbor Ising interaction, a smaller XY exchange $\alpha_\text{S}$, and a bond-dependent staggered YZ interaction $\alpha_{yz}$, along with the transverse field term. The bond dependence of the staggered YZ term arises from the zig-zag nature of the chains. The mean-field term $\mathcal{H}_3$ accounts for the effective magnetic field on a given chain arising from weak, frustrated antiferromagnetic interchain couplings, which becomes relevant only below the ordering temperature $T_\mathrm{N}$. In what follows, we set $\braket{S^y}=0$ and $\braket{S^z}=0.5$, independent of the applied field strength (an excellent approximation for the low-field regime we will be interested in). Finally, an additional intra-chain $\mathcal{H}_2$ includes only small corrections, which are known to further refine the quantitative agreement with inelastic neutron scattering data \cite{Woodland2023}.

\section{Review of 2DCS and methods} 
\label{sec:methods}

\begin{figure}
	\centering
	\includegraphics[width=1.0\columnwidth]{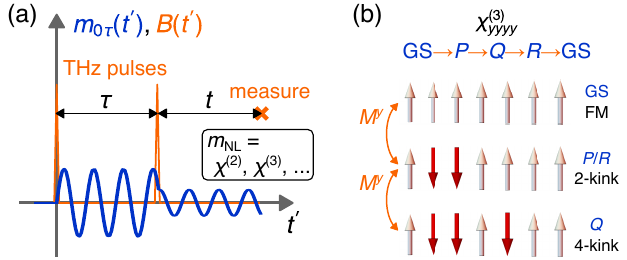}
	\caption{{\bf Probing the 4-kink sector with nonlinear spectroscopy.} (a) Schematic of a two-pulse THz magnetic-field protocol used to measure the nonlinear magnetization response. The nonlinear magnetization $m_{\mathrm{NL}}$ contains contributions from susceptibilities $\chi^{(n)}$ with $n\ge 2$. (b) An example excitation pathway driven by $M^y$, involving the ground state (GS) as the initial and final states and intermediate states $\ket{P}, \ket{Q}, \ket{R}$ that contribute to the third-order susceptibility $\chi^{(3)}_{yyyy}$. The third-order process involves four insertions of $M^y$ (three from the field pulses and one at detection), indicated by orange arrows. Spins are shown schematically as arrows (white/red denote up/down). Starting from the fully polarized ferromagnetic (FM) state, one application of $M^y$ can create a reversed-spin domain bounded by two kinks (two spinons). A second application can reach four-kink intermediate states $\ket{Q}$.}
	\label{fig:tdcs_schematics}
\end{figure}

Let us start by briefly reviewing the 2DCS set-up for our case---for a general discussion we refer to Ref.~\cite{Watanabe2024}. We apply a static magnetic field along the crystallographic $b$ axis (the $y$ direction in the local spin frame), as well as two magnetic-field pulses, focusing here only on the case in which the pulses are aligned along the same crystallographic direction. The two magnetic-field pulses are separated by a time delay $\tau$, and the magnetization is measured a time $t$ after the second pulse (i.e.~at a total time $t+\tau$ after the first pulse). For simplicity, we model the pulses as delta functions in time,
\[
B^y(t') = B_0^y \,\delta(t') + B_\tau^y \,\delta(t'-\tau),
\]
where $B_0^y$ and $B_\tau^y$ are the pulse amplitudes.

Let $m_{0\tau}^y(t,\tau)$ denote the induced $y$-component of the magnetization after the two pulses. Similarly, $m_0^y(t,\tau)$ and $m_\tau^y(t,\tau)$ denote the induced responses to the first or second pulse alone (with the other pulse absent). Subtracting the single-pulse responses from the two-pulse response isolates the nonlinear magnetization,
\begin{equation}
    \begin{split}
         m_{\text{NL}}^y(t,\tau) =&\,\, m^y_{0\tau}(t,\tau) - [m^y_{0}(t,\tau) + m^y_{\tau}(t,\tau)] \\
         =
				 &\,\, \chi_{yyy}^{(2)}(t, \tau)B^y_\tau B_0^y\\
				 & + \chi_{yyyy}^{(3)}(t, \tau, 0)B_\tau^y (B_0^y)^2\\
				 & + \chi_{yyyy}^{(3)}(t, 0, \tau) (B_\tau^y)^2 B_0^y\\
				 & + \text{(higher-order contributions)},
    \end{split}
    \label{eq:M_NL}
\end{equation}
so that only mixed cross-terms survive. This set-up is illustrated schematically in Fig.~\ref{fig:tdcs_schematics}.

Among the experimentally accessible nonlinear susceptibilities, we focus on $\chi^{(3)}_{yyyy}(t,0,\tau)$, which can be written in Lehmann representation as~\cite{Watanabe2024}
\begin{equation}
 \begin{split}
    \chi^{(3)}_{yyyy}(t,0,\tau) &= -\frac{1}{N}\sum_{PQR}
       m^y_{0P} m^y_{PQ} m^y_{QR} m^y_{R0} \\
    &\quad \times \Big[ \sin(\omega_P\tau+\Delta\omega_{PQ}t) \\
    &\qquad + 2\sin(-\omega_R\tau+\Delta\omega_{PQ}t) \\
    &\qquad + \sin(\omega_P\tau+\omega_R t) \Big] \,,
 \end{split}
 \label{eq:chi3}
\end{equation}
where $m^y_{ab}=\langle a|M^y|b\rangle$, $\Delta\omega_{ab}=\omega_a-\omega_b$, and $M^y=\sum_j S_j^y$ is the total $y$-magnetization. We analyze the imaginary part of the two-dimensional Fourier transform with respect to the two times $t$ and $\tau$
\begin{equation*}
\mathrm{Im}\left[\chi^{(3;2)}_{yyyy}(\omega_t, \omega_\tau)\right] = \mathrm{Im}\left\{\mathcal{FT}_{t,\tau}\left[\chi^{(3)}_{yyyy}(t,0,\tau)\right]\right\} \,,
\end{equation*}
which is the main quantity computed and discussed below (the superscript (3;2) indicates that the time between the second and third insertions of $M^y$, often referred to as $t_2$, is set to zero). In the experiment, both $t$ and $\tau$ are constrained to be positive, $t, \tau > 0$. It is known that this restriction leads to a {\it phase-twisting} effect that distorts certain features in the two-dimensional frequency plane~\cite{Watanabe2024}. For analysis purposes, we occasionally extend the $\tau$ integral to negative values to remove this distortion, yielding what we refer to as the {\it phase-untwisted} spectra.

As expressed in Eq.~(\ref{eq:chi3}), the third-order response involves three intermediate states, $\ket{P}$, $\ket{Q}$, and $\ket{R}$, with the ground state $\ket{0}$ serving as both the initial and final state. When certain symmetries are present, the Hilbert space splits into sectors labeled by the associated conserved quantum numbers (e.g., $S_z$, momentum, parity). Suppose the system couples to a probe operator $\mathcal{O}$ (here, $M^y$) that carries a definite symmetry charge. In linear response,
\[
\chi^{(1)}(\omega)\propto \sum_P \bigl|\braket{P|\mathcal{O}|0}\bigr|^2\,\delta(\omega-\omega_P),
\]
only excited states $\ket{P}$ with $\braket{P|\mathcal{O}|0}\ne 0$ are accessible, as set by selection rules. In contrast, the third-order response contains terms where $\mathcal{O}$ acts twice in succession, which can reach intermediate states $\ket{Q}$ that are symmetry-forbidden in linear response. This provides access to otherwise hidden symmetry sectors~\cite{Watanabe2025}. Throughout this work, we identify peaks in the 2DCS spectra as signatures of such $\ket{Q}$ intermediate states (rather than $\ket{P}$ or $\ket{R}$).

Note that in the absence of symmetry constraints, most excited states generically have finite matrix elements with the ground state, so linear response can in principle resolve the full excitation spectrum. In that case, the benefit of nonlinear spectroscopy lies not in accessing forbidden transitions, but in enhancing the visibility of weakly coupled states and in probing interaction effects between excitations, including statistical correlations~\cite{Fava2020,McGinley2024}.

\subsection*{Four-kink approximation}
The ability to access higher-order excitations in 2DCS is a clear advantage, but it also complicates the required theory, i.e., approximations that neglect higher-order or multiparticle excitations, which are often justified in linear-response calculations, may fail to capture, even qualitatively, the nonlinear spectra. In the present case, a single application of $M^y$ connects the fully polarized ferromagnetic ground state primarily to states with two kinks (two spinons). Therefore, the linear response of \CoNbO\  in zero and small external fields, away from the critical point, is well captured by restricting the Hilbert space to the two-kink subspace~\cite{Woodland2023}. However, in the third-order response, two applications of $M^y$ can naturally create up to four kinks/spinons [Fig.~\ref{fig:tdcs_schematics}(b)].

Accordingly, we extend the standard two-kink approximation for \CoNbO\ to include the four-kink subspace when computing 2DCS signatures. We truncate the Hilbert space around the fully polarized ferromagnetic state to sectors with up to four kinks. Inter-sector couplings are eliminated, i.e., we perform a Schrieffer-Wolff transformation to leading order, yielding a kink-number-conserving effective Hamiltonian \cite{Woodland2023}. In our calculation, couplings between different kink-number sectors enter only through the external THz pulses. Including up to four-kink states has previously been shown to be necessary to capture key features of the 2DCS spectra of the 1$d$-TFIM in a small longitudinal field~\cite{Watanabe2024}.

Each two-kink state is labeled by two quantum numbers, $\ket{i, l}$, where $i$ denotes the position of the left kink and $l$ specifies the length of the domain attached to it. A four-kink state is then labeled by $\ket{i_1, l_1, i_2, l_2}$, corresponding to two such domain-wall pairs. As a result, the Hilbert space scales as $\mathcal{O}(L^4)$, where $L$ is the number of sites. By working in the $k = 0$ momentum sector, the number of independent quantum numbers can be reduced to three, allowing simulations of systems with up to $\mathcal{O}(100)$ sites. It has been shown that, in the presence of a confining potential that discretizes the spectrum, finite-size effects are minor, and the small system sizes accessible to exact diagonalization (ED) are sufficient to capture the relevant physics~\cite{Watanabe2024}. For our model, if we compare ED calculations for  system size $L = 20$ to four-kink calculations for the same system size, as shown in Appendix~\ref{app:ED_vs_4kink}, we find almost perfect qualitative agreement (up to a rescaling of the magnetic field strength by a factor of $2/3$).
In the main text we focus on the results of numerical four-kink calculations, and throughout this work we fix the system size to $L=100$ sites.

In addition to computing the 2DCS signal itself, we also evaluate the dynamical spin structure factor at zero momentum
\[
S(\omega) = \mathcal{FT}\,[\langle M_y(t)M_y(0)\rangle]
\]
and its higher-order counterpart
\[
I(\omega) = \mathcal{FT}\,[\langle M_y^{2}(t)M_y^{2}(0)\rangle] \,.
\]
We plot $S(\omega)$ on a linear scale and $I(\omega)$ on a logarithmic scale, since in $I(\omega)$ the spectral weight from the two-kink sector (around $2.5~\mathrm{meV} \approx J$) is much smaller than that from the four-kink sector (around $5~\mathrm{meV} \approx 2J$). Although $I(\omega)$ is not directly measurable in experiments, it can, in principle, be reconstructed from the full 2DCS signal. Here, we take the reverse approach and use $I(\omega)$ as a diagnostic tool to aid the interpretation of the 2DCS spectra.

It is important to note that the relative spectral weights of the two-kink and four-kink sectors in $I(\omega)$ do not, by themselves, imply that four-kink contributions dominate the 2DCS response. Different excitation pathways---particularly those involving intermediate states $Q$ in the ground state (0-kink) sector or in the four-kink sector---can contribute with opposite signs at the same locations in the two-dimensional frequency plane~\cite{Watanabe2024}. The 2DCS spectra thus reflect the residual signal after such interference and cancellation effects.

\section{Step-by-step construction of spectroscopic signatures}
\label{sec:construction}

\begin{figure}[b]
	\centering
	\includegraphics[width=0.95\columnwidth]{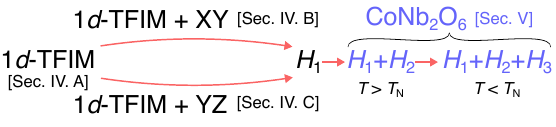}
	\caption{{\bf Construction of the full Hamiltonian.} Starting from the exactly solvable 1$d$-TFIM Hamiltonian, we gradually introduce additional interaction terms present in \CoNbO. The most dominant terms are included in $\mathcal{H}_1$, which consists of the Ising interaction along with subdominant XY and YZ terms. $\mathcal{H}_2$ is introduced to achieve quantitative agreement with inelastic neutron scattering experiments. The interchain coupling is represented by the mean-field term $\mathcal{H}_3$, which becomes relevant only below $T_\mathrm{N}$.}
	\label{fig:interpolations}
\end{figure}

In the following we will proceed with a step-by-step approach to construct the 2DCS signatures for the full Hamiltonian of \CoNbO, starting first from the exactly solvable 1$d$-TFIM and successively adding the additional interaction terms present in \CoNbO. This procedure is schematically illustrated in Fig.~\ref{fig:interpolations}.
We start, in this section, by focusing on $\mathcal{H}_1$. We first present the 2DCS spectra of the 1$d$-TFIM and use them to introduce basic features of 2DCS in 1$d$ models with fractionalized spinon excitations, along with standard terminology for labeling signals. We then add the additional interaction terms included in $\mathcal{H}_1$ (the XY and YZ terms), which are comparable in strength and are the second-largest exchange terms after the Ising term---we add them separately to isolate their individual effects. We find that the additional YZ term is the most relevant for understanding the 2DCS spectra of \CoNbO\ within our low-field parameter regime. We also show that the other subdominant term, the XY interaction, plays a comparatively minor role for the known parameters of \CoNbO. While the XY term can, in general, stabilize a single-spin-flip bound state by pulling it out of the continuum, this stabilization is not realized here; instead, we mainly observe the associated precursor renormalization of the continuum and a net shift of spectral weight to lower energies, resulting in an overall spectral shift without qualitatively changing the 2DCS features. After this, in Sec.~\ref{sec:full_spectrum}, we will present the 2DCS spectra of the full Hamiltonian and trace their evolution along an interpolation from the TFIM+YZ model all the way to the final full Hamiltonian.

\subsection{1$\mathbf{d}$ Transverse-field Ising model}

\begin{figure}[t]
	\centering
	\includegraphics[width=1.0\columnwidth]{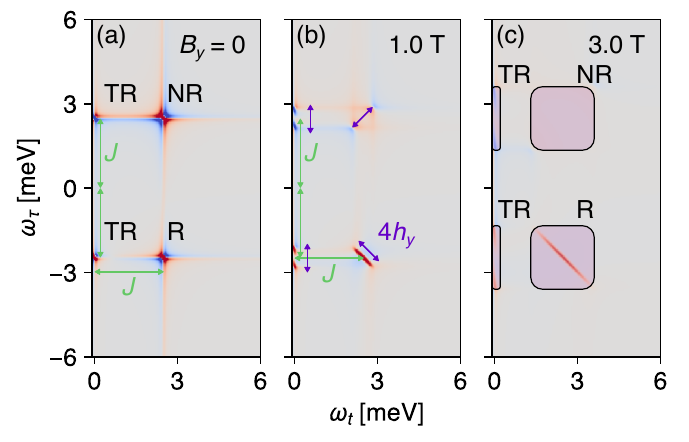}
	\caption{{\bf  2DCS spectra of transverse-field Ising chain.} Imaginary part of third-order response $\mathrm{Im}[\chi^{(3;2)}_{yyyy}(\omega_t, \omega_\tau)]$ for a chain with $L=100$ sites at transverse fields $B_y = 0,\, 1$, and $3$~T, shown from (a) to (c). At $B_y = 0$, all non-rephasing (NR), rephasing (R), and terahertz-rectification (TR) signals appear as point-like features. A finite $B_y$ introduces dispersion to the spinons, leading to a finite energy bandwidth of $4h_y = 4g\mu_\mathrm{B}B_y$ centered at $J$. As a result, all point signals stretch into continuous line structures, as indicated by the purple arrows, with the NR feature barely visible due to the phase-twisting effect.}
	\label{fig:tfim}
\end{figure}

First, we consider the effect of the dominant FM Ising interaction together with the transverse field, i.e., the 1$d$-TFIM. The model is exactly solvable via a Jordan-Wigner transformation, and its 2DCS spectra have been studied in detail in Ref.~\cite{Wan2019}, which reported sharp ``spinon-echo'' signals in the third-order response. Here we briefly summarize the key features.

In the absence of the transverse field, $h_y = 0$, the system reduces to the 1$d$ classical FM Ising model, where every spin configuration in the $z$ basis is an eigenstate, implying trivial classical dynamics. Each $S^y_i$ operator flips a single spin at site $i$, costing an energy $J$ and creating two kinks on both sides of the flipped spin.

Consequently, the intermediate $\ket{P}$ and $\ket{R}$ states can only correspond to single spin-flip (SF) excitations with $\omega_{P/R} = J$. For the $\ket{Q}$ state, it can either be the ground state, a two-adjacent-SF state with $\omega_Q = J$, or a two-separated-SF state with $\omega_Q = 2J$ (alternatively, if one considers the single-SF as a single magnon, then these two-spin-flip states would be a two-magnon bound state and two separated one-magnon states respectively). Inserting this information into Eq.~(\ref{eq:chi3}) with the appropriate counting of each process yields the spectrum shown in Fig.~\ref{fig:tfim}(a) (the figure itself is obtained numerically via the four-kink calculation). Using the standard terminology for non-linear response signals, we observe sharp non-rephasing (NR), rephasing (R), and terahertz-rectification (TR) signals as point-like features at $(\omega_t, \omega_\tau) = (J, J)$, $(J, -J)$, and $(0, \pm J)$, respectively.

When $h_y$ is finite, the spinon excitations become dispersive. The operator $M^y$ can be written as $M^y = \sum_i S_i^y = \sum_k m_k$, where each $m_k$ creates a spinon pair with momenta $k$ and $-k$~\cite{Wan2019}. The intermediate $\ket{P}$ and $\ket{R}$ states correspond either to the ground state or to a two-spinon state with energy $\omega_{P/R} = \sqrt{J^2 + 4h_y^2 - 4Jh_y\cos(k)}$, which spans the range $[J - 2h_y, J + 2h_y]$. The $\ket{Q}$ state can be the ground state, a two-spinon state, or a four-spinon state composed of $\pm k$ and $\pm k'$ pairs. In this particular case, however, since $M^y$ can be decomposed into mutually commuting operators $m_k$, we only need to consider processes in which every spinon pair appearing in $\ket{P}$, $\ket{Q}$, and $\ket{R}$ carries the same momentum $k$, and the final response is obtained by summing over $k$.

As a result, the previously point-like R and TR signals are stretched into continuous line features, as shown in Figs.~\ref{fig:tfim}(b, c), with each point along the line corresponding to a particular momentum $k$. The NR signal extends along the diagonal direction, but such continuous diagonal features are suppressed by the phase-twisting effect~\cite{Watanabe2024}, making them barely visible in the figure.

The R signal, stretched along the anti-diagonal direction $\omega_t = -\omega_\tau$---the so-called spinon echo, whose signature appears at $t = \tau$ in the time domain---has attracted considerable attention. The sharpness of this feature along the diagonal direction reflects the infinite lifetime of spinons in the 1$d$-TFIM; if interactions and decay were present, one would expect this feature to broaden along the diagonal direction depending on the lifetime, thereby providing a means to probe the nature of fractionalized excitations~\cite{Wan2019,Hart2023}. We will revisit this point in the next section, where the additional XY term introduces interactions between spinons.

\subsection{Ising + XY + Transverse Field}
\label{sec:ising_xy}

Next we investigate the effect of adding the XY term to the 1$d$-TFIM, yielding the ferromagnetic XXZ model in a uniform transverse magnetic field. The linear-response signatures of this model are summarized in Fig.~\ref{fig:xy}.
The 2DCS response of this model was analyzed in Ref.~\cite{Hart2023} using a diagrammatic approach, where the XY term was treated perturbatively starting from the pure TFIM. Within that framework, the XY interaction---although introducing spinon-spinon interactions---does not generate a finite spinon lifetime at zero temperature, since a single spinon lacks sufficient phase space to decay into the three-spinon continuum. Consequently, the spinon-echo line remains sharp, with the XY term primarily shifting spectral weight toward lower frequencies.

Our numerical approach, which does not rely on a perturbative expansion about the TFIM limit, reveals additional features that are not captured by the diagrammatic treatment in Ref.~\cite{Hart2023}. For example, our calculation naturally includes processes in which $\ket{P}$ and $\ket{R}$ correspond to spinon pairs with different momenta, such as $(-k_P, k_P)$ and $(-k_R, k_R)$ with $k_P \neq k_R$. These momentum-mismatched processes are not captured in the perturbative framework, which focuses on renormalizing the single-particle propagator, but they give rise to additional off-diagonal features away from the spinon-echo line in the full 2DCS response. In the following, we first analyze the tractable $h_y = 0$ case and then present results for finite $h_y$, where these effects begin to manifest in the numerical spectra.

\subsubsection*{$h_y = 0$ case}
\begin{figure}
	\centering
	\includegraphics[width=1.0\columnwidth]{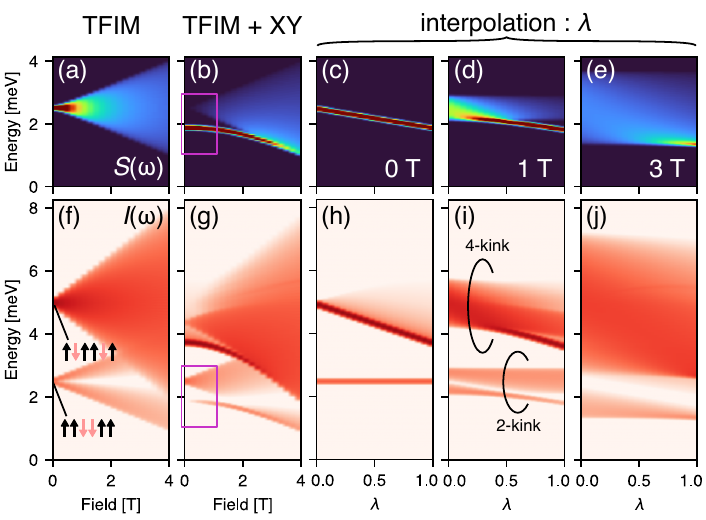}
	\caption{{\bf Linear response spectra of TFIM + XY}. (a--e) Dynamical structure factor $S(\omega)=\mathcal{FT}\,[\langle M_y(t)M_y(0)\rangle]$ and (f--j) its higher-order counterpart $I(\omega)=\mathcal{FT}\,[\langle M_y^{2}(t)M_y^{2}(0)\rangle]$. Linear scales are shown in panels (a--e), and logarithmic scales in panels (f--j). Both spectra were obtained using a four-kink calculation with $L = 100$ sites. (a, f) Transverse-field dependence of the Ising-model spectra. (b, g) Transverse-field dependence of the Ising + XY model. Purple boxes highlight the transverse-field-induced hybridization of the (confined) single spin-flip state with the spinon continuum. (c--e, h--j) Interpolation between the pure Ising limit and the Ising model with additional XY terms ($\alpha_{\mathrm{S}} = 0.251$) for $B_y = 0,\, 1,\, \text{and}\, 3~\text{T}$, parameterised by $\lambda \in [0,1]$.}
	\label{fig:xy}
\end{figure}

When $h_y = 0$, the Ising+XY Hamiltonian possesses a $U(1)$ symmetry---meaning that total $M^z$ is conserved. The operator $M^y$ changes $\abs{M^z}$ by 1, so the intermediate states $\ket{P}$ and $\ket{R}$ in Eq.~(\ref{eq:chi3}) must belong to the single-SF sector, which forms a subspace of the two-kink sector. From the spinon perspective, the XY term enables the hopping of a spinon pair when the two spinons are adjacent, thereby introducing dispersion to the single-SF mode (most naturally seen by writing the XY term in terms of $S_j^+ S_{j+1}^- + S_j^- S_{j+1}^+$). The hopping amplitude is $J\alpha_\mathrm{S}/2$, resulting in a bandwidth of $2J\alpha_\mathrm{S}$ and an energy range $[J(1 - \lambda\alpha_\mathrm{S}),\, J(1 + \lambda\alpha_\mathrm{S})]$, where $\lambda \in [0,1]$ serves as an interpolation parameter: $\lambda = 0$ corresponds to the pure Ising limit, and $\lambda = 1$ to the full XY strength of \CoNbO. Since $M^y$ creates single-SF excitations only at $k = 0$, the spectral weight in $S(\omega)$ is localized exactly at the lower edge of the single-SF dispersion, i.e., at $\omega = J(1 - \lambda\alpha_\mathrm{S})$ [Fig.~\ref{fig:xy}(c)].

Applying $M^y$ twice accesses the $\abs{\Delta M^z} = 2$ sector, corresponding to two-spin-flip (2SF) excitations. Among these, the two-adjacent-SF state belongs to the two-kink sector, while the two-separated-SF states belong to the four-kink sector. Within our kink-number-conserving approximation, the XY term does not induce dynamics for the two-adjacent-SF state---its action on it would generate a four-kink configuration, which is projected out---so this mode remains dispersionless (localized) at energy $J$, independent of $\lambda$ [Fig.~\ref{fig:xy}(h)]. This localized mode is referred to as the 2SF bound state.

In contrast, the two-separated-SF states can be viewed as two nearly independent single-SF excitations, each acquiring dispersion from the XY term and spanning the energy range $[2J(1 - \lambda\alpha_\mathrm{S}),\, 2J(1 + \lambda\alpha_\mathrm{S})]$---the \textit{2SF continuum}. Since two single-SF excitations can be created with zero total momentum, the spectrum $I(\omega)$ thus exhibits a continuum extending from $2J(1 - \lambda\alpha_\mathrm{S})$ to $2J(1 + \lambda\alpha_\mathrm{S})$ [Fig.~\ref{fig:xy}(h)] with spectral weight strongly concentrated near the lower edge $2J(1 - \lambda\alpha_\mathrm{S})$ and a finite tail persisting toward higher energies.

\subsubsection*{$h_y \ne 0$ case}
Finite $h_y$ breaks the $U(1)$ symmetry, generating hybridization between different total $M^z$ sectors, particularly between the single-SF ($\abs{\Delta M^z} = 1$) and two-adjacent-SF ($\abs{\Delta M^z} = 2$) excitations, which are further dynamically connected to higher SF sectors (though in our numerics restricted to the same kink number). Consequently, $S(\omega)$ acquires a finite contribution from the two-adjacent-SF states, forming a continuum due to the spinon dispersion introduced by $h_y$. Conversely, $I(\omega)$ gains spectral weight from the single-SF bound states through hybridization. In weak fields, the single-SF energy remains well separated from the two-adjacent-SF continuum, so the hybridization effect is small. As $h_y$ increases, the single-SF energy approaches the continuum, enhancing the hybridization and transferring more spectral weight [Figs.~\ref{fig:xy}(e) and \ref{fig:xy}(j), highlighted by the purple boxes]. Once $h_y$ becomes sufficiently large, the single-SF mode merges into the continuum (in our few-kink calculation, this occurs around $B_y \approx 2.5$~T).

To view this from another perspective, when taking the 1$d$-TFIM with finite $h_y$ as a starting point, the XY term introduces interactions between spinons, which renormalize the continuum and enhance the spectral weight near its lower edge [Figs.~\ref{fig:xy}(d) and \ref{fig:xy}(e)]. Once the XY term becomes sufficiently strong, the single-SF bound state detaches from the continuum---a feature visible at 1~T around $\lambda \approx 0.5$, but only barely avoided at 3~T for $\lambda = 1.0$. As we will see later, in \CoNbO, where an additional YZ interaction is present and effectively acts as a staggered field, the role of the XY term becomes comparatively minor, resulting mainly in a spectral shift.

\begin{figure}[t]
	\centering
	\includegraphics[width=1.0\columnwidth]{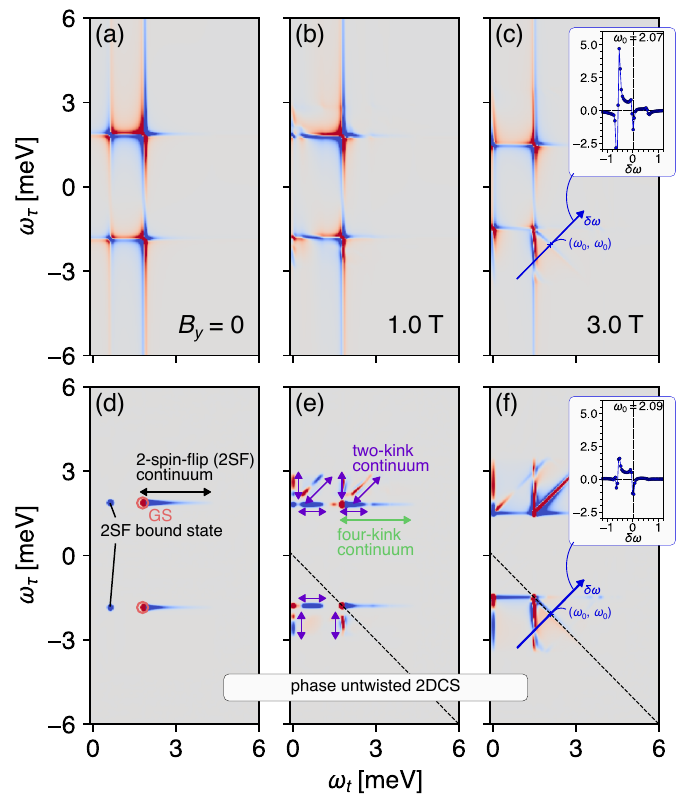}
	\caption{{\bf 2DCS spectra of TFIM + XY.} (a--c): experimentally accessible signals for each magnetic field. (d--f): corresponding phase-untwisted spectra. At $B_y = 0$, the intermediate states $\ket{P}$ and $\ket{R}$ are restricted to single spin-flip (SF) states. $\ket{Q}$ can be either the ground state (GS), a two-adjacent-SF bound state (2-kink sector), or a two-separated-SF state (4-kink sector), the latter forming a continuum. A finite $B_y$ induces hybridization within both the 2-kink and 4-kink sectors, leading to additional elongated features (marked by purple arrows) corresponding to the two-kink continuum; $\ket{P}$ and $\ket{R}$ can now be either a 1-SF bound state or part of the 2-kink continuum. The phase-twisting effect suppresses the diagonal line features in (a--c). For $B_y = 3.0$~T, a diagonal line cut at $(\omega_t, \omega_\tau) = (\omega_0 + \delta\omega, \omega_0 + \delta\omega)$ is shown in the insets of (c) and (f). The anti-diagonal R signal at $\omega_t = -\omega_\tau$ shows a sharp peak along the cut accompanied by a sign-changing broad background.}
	\label{fig:tfim_xx_yy}
\end{figure}

\subsubsection*{2DCS spectra}

Let us now turn to the 2DCS spectra of the TFIM+XY model ($\lambda = 1$), shown in Fig.~\ref{fig:tfim_xx_yy}. At $h_y = 0$ ($B_y = 0$), the intermediate states $\ket{P}$ and $\ket{R}$ are restricted to single-SF states at $k = 0$ with energy $J(1 - \alpha_\mathrm{S})$, while the $\ket{Q}$ state can be either the ground state, the 2SF bound state, or the 2SF continuum, as discussed above. The spectra in Figs.~\ref{fig:tfim_xx_yy}(a) and \ref{fig:tfim_xx_yy}(d) can be understood by inserting these energy levels into Eq.~(\ref{eq:chi3}). For example, the two peaks appearing at $(\omega_t, \omega_\tau) = [J\alpha_\mathrm{S},\, J(1 - \alpha_\mathrm{S})] \approx (0.62, 1.86)~\text{meV}$ originate from the pathway where $\ket{Q}$ corresponds to the 2SF bound state. The NR/R peaks at $(\omega_t, \omega_\tau) = [J(1 - \alpha_\mathrm{S}),\, \pm J(1 - \alpha_\mathrm{S})]$ arise from two competing contributions---one from the process $\ket{Q} = \ket{0}$ and the other from the lower edge of the 2SF continuum---and the observed signal represents the residual after their partial cancellation. The streak-like feature emanating from the R/NR peaks along the horizontal direction is due to the 2SF continuum component of $\ket{Q}$ and is particularly prominent in the phase-untwisted spectrum [Fig.~\ref{fig:tfim_xx_yy}(d)].

For small $h_y$ (e.g., $B_y = 1.0$~T), the intermediate states $\ket{P}$ and $\ket{R}$ can either be the single-SF bound state or part of the two-kink continuum, and $\ket{P}$ and $\ket{Q}$ do not necessarily coincide. The $\ket{Q}$ state can be the single-SF bound state, the two-kink continuum, or the four-kink continuum. As a result, compared to the $B_y = 0$ case, the spectra exhibit peak splitting due to hybridization and additional line-like features, some of which are indicated by colored arrows in Fig.~\ref{fig:tfim_xx_yy}(e). In the experimental signal, diagonal features are again suppressed by the phase-twisting effect, as shown in Fig.~\ref{fig:tfim_xx_yy}(b).

At larger $h_y$ (e.g., $B_y = 3.0$~T), the single-SF enters the continuum. From the viewpoint of the 1$d$-TFIM, this corresponds to a regime where the model hosts a continuum similar to that of the 1$d$-TFIM, but renormalized by spinon interactions introduced by the XY term. It is therefore instructive to examine how the spinon-echo spectrum along $\omega_t = -\omega_\tau$ behaves---this model serves as a useful case study of interacting fractionalized excitations distinct from the lifetime effects arising from the self-energy correction of individual spinons. Indeed, the XY interaction itself does not induce a finite spinon lifetime~\cite{Hart2023}; thus, if only 
a finite-lifetime broadening mechanism were to be considered, no additional features off of the anti-diagonal line would be expected in the 2DCS spectrum. However, as shown in Figs.~\ref{fig:tfim_xx_yy}(c) and \ref{fig:tfim_xx_yy}(f) and their insets, our four-kink calculation reveals additional features off of the anti-diagonal, most prominently a vertical cross-feature, together with a relatively sharp signal centered along the spinon-echo line. The sign of the spinon-echo peak is now negative (whereas it was positive in the 1$d$-TFIM case), suggesting that a sign reversal occurs between $\lambda = 0$ (1$d$-TFIM) and $\lambda = 1$ (TFIM+XY). This points to a qualitative modification of the 2DCS spinon-echo response driven by spinon interactions beyond a simple lifetime broadening mechanism. These observations call for a re-examination of theoretical predictions for interacting spinons, as 2DCS may provide a sensitive probe of such interaction effects---though the extent of this capability remains to be established in future work.

\subsection{Ising + Staggered YZ + Transverse Field}
\label{sec:ising_yz}

\begin{figure}[t]
	\centering
	\includegraphics[width=1.0\columnwidth]{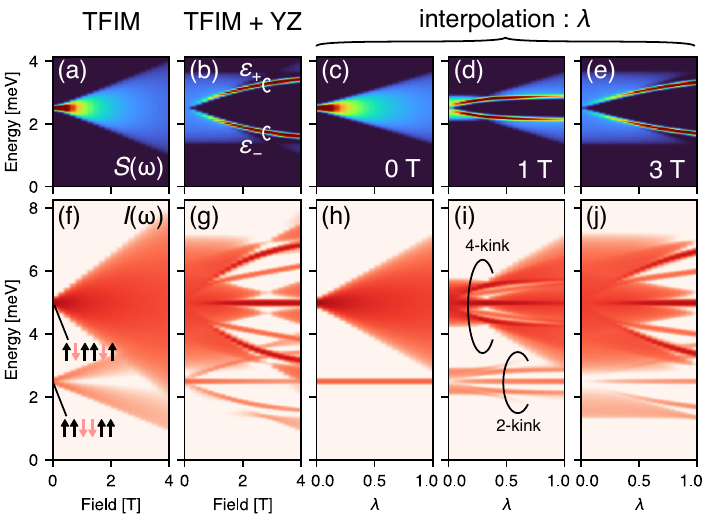}
	\caption{{\bf Linear response spectra of TFIM + YZ}. (a--e) Dynamical structure factor $S(\omega)$ and (f--j) its higher-order counterpart $I(\omega)$. Linear scales are shown in panels (a--e), and logarithmic scales in panels (f--j). Both spectra were obtained using a four-spinon calculation with $L = 100$ sites. (a, f) Transverse-field dependence of the Ising-model spectra. (b, g) Transverse-field dependence of the Ising + YZ model. (c--e, h--j) Interpolation between the pure Ising limit and the Ising model with additional YZ terms ($\alpha_{\mathrm{yz}} = 0.226$) for $B_y = 0,\, 1,\, \text{and}\, 3~\text{T}$, parameterized by $\lambda \in [0,1]$.}
	\label{fig:yz}
\end{figure}

\begin{figure*}[t]
	\centering
	\includegraphics[width=1.0\textwidth]{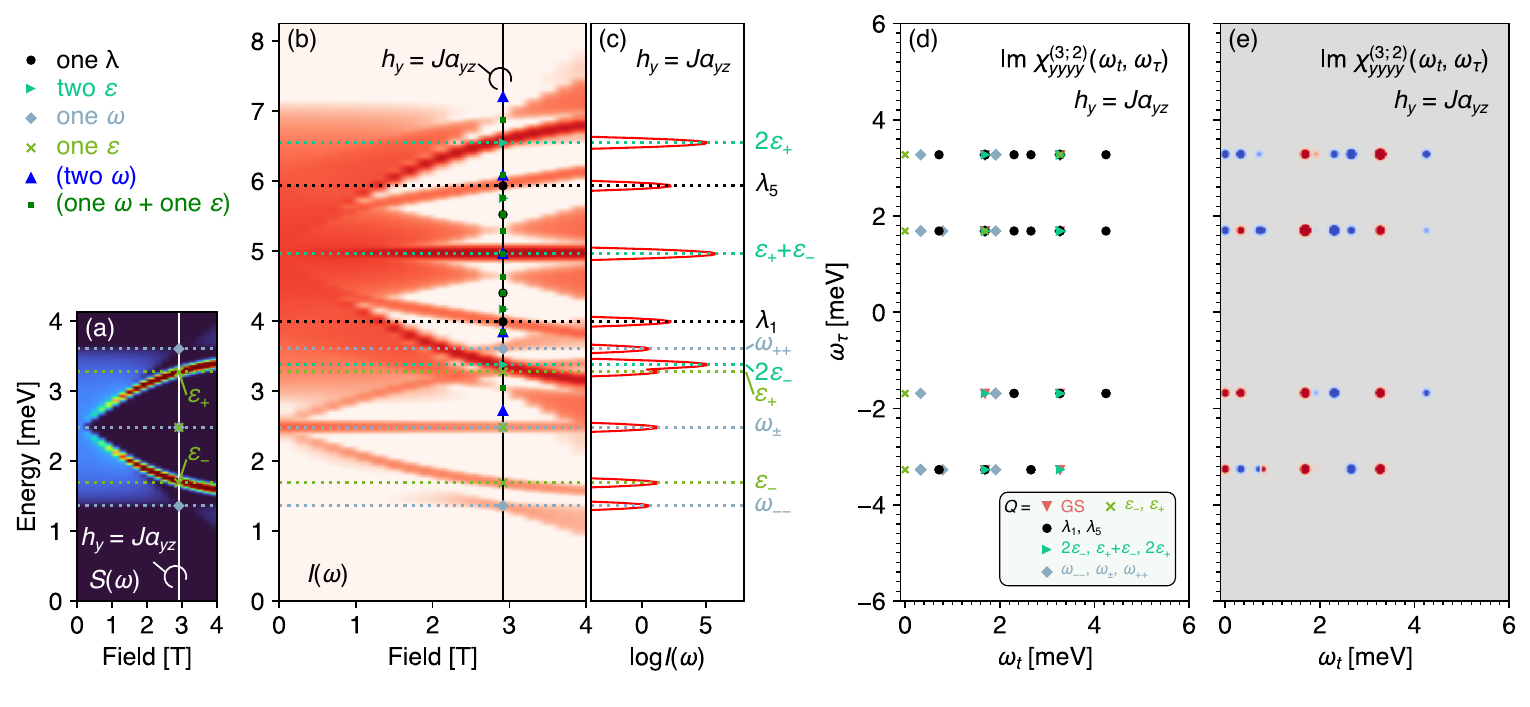}
	\caption{\textbf{Localized limit}.
	(a) Field dependence of $S(\omega)$. The energies of the two-kink modes at the localized limit, defined by $h_y = J\alpha_{yz}$, are indicated by dashed lines; among these, only the $\varepsilon_{\pm}$ modes are visible in $S(\omega)$.
	(b) Field dependence of $I(\omega)$. The energies of the two- and four-kink modes at the localized limit are marked with distinct symbols according to the mode type. Dotted lines serve as eye guides highlighting the modes that remain visible at the localized limit.
	(c) Line cut of $I(\omega)$ at $h_y = J\alpha_{yz}$.
	(d) Schematic 2DCS spectrum at the localized limit. Peaks are distinguished by different symbols depending on the intermediate state $\ket{Q}$ in the pathway $0 \rightarrow P \rightarrow Q \rightarrow R \rightarrow 0$, where $P$ and $R$ are always either $\varepsilon_{+}$ or $\varepsilon_{-}$. (e) Numerically obtained 2DCS spectrum at the localized limit.}
	\label{fig:tfim_yz_detail}
\end{figure*}

We now discuss the effect of the other subdominant interaction in \CoNbO, namely, the staggered YZ term, which takes the form 
\[
	\sum_j (-1)^j \alpha_{yz} \left( S_j^y S_{j+1}^z + S_j^z S_{j+1}^y \right)
\]
with $\alpha_{yz} = 0.226$. Although slightly smaller in magnitude than the XY term, it turns out that many of the key 2DCS features discussed in this section carry over directly to the case of \CoNbO.

The YZ term behaves similarly to a staggered transverse field, introducing dispersion to the spinon excitations. This can be seen by considering the combined action of the transverse field term and the YZ term on a single kink (where we fix $\lambda = 1.0$). As an example, take a left kink state $\ket{\psi}_\mathrm{L} = \sum_{j'} \psi_{\mathrm{L}}(j')\ket{j'}_\mathrm{L}$ with energy $\omega$, where $\ket{j} = \ket{\cdots \uparrow\uparrow\uparrow_{j-1} \downarrow_j \downarrow \downarrow \cdots}$. The Schrödinger equation for the kink wavefunction $\psi_\mathrm{L}(j)$ \cite{Woodland2023} reads
\begin{equation}
\left(\omega - \frac{J}{2}\right)\psi_\mathrm{L}(j) = \frac{1}{2}\sum_{\Delta = \pm 1}\left[h_y + (-1)^j J\alpha_{yz}\Delta\right]\psi_\mathrm{L}(j - \Delta),
\end{equation}
representing staggered hopping of the kink with hopping amplitudes $h_\pm = (1/2)(h_y \pm J\alpha_{yz})$.


When $h_y=0$, the Hamiltonian of this two-sublattice model resembles that of the twisted Kitaev chain~\cite{Morris2021}, which hosts non-interacting spinon excitations. Indeed, the model can be solved exactly via a Jordan-Wigner transformation, with the resulting two fermionic bands having energies $\xi^{\pm}_k = (J/2)[1 \pm 2\alpha_{yz}\cos(k)]$. The two bands lead to three different types of spinon pair excitations, two intra-band pair excitations and one inter-band pair excitation. For the present set of parameters, the spinon continua arising from these distinct pair excitations all overlap in energy. As a result, the linear response $S(\omega)$ and $I(\omega)$ exhibit a continuum even without having explicit transverse field $h_y$, as shown in Figs.~\ref{fig:yz}(c) and \ref{fig:yz}(h).

\subsubsection*{Localized limit}

Within the four-kink approximation, the model admits another solvable point, namely 
\[
	h_y = J\alpha_{yz} \,,
\] where $h_{-}$ vanishes. 
At this point, the left kink is only allowed to move between $\ket{2p-1}_\mathrm{L}$ and $\ket{2p}_\mathrm{L}$, whereas the right kink is only allowed to move between $\ket{2p}_\mathrm{R}$ and $\ket{2p+1}_\mathrm{R}$, each with a hopping amplitude $h_{+} = J\alpha_{yz}$. Here, $p \in \intg$ is an index specifying a unit cell rather than a site. This restricts the dynamics of the kinks, and defines the so-called {\it localized limit}, introduced and discussed in detail within the two-kink sector in Ref.~\cite{Woodland2023} and which we briefly recap here. When the two kinks are far apart, the dynamics of the left and right kinks decouple. For the left and right kinks, the eigenstates are given by $\ket{p}^{\pm}_\mathrm{L} = (1/\sqrt{2}) \left(\ket{2p-1} \pm \ket{2p}\right)$ and $\ket{p}^{\pm}_\mathrm{R} = (1/\sqrt{2}) \left(\ket{2p} \pm \ket{2p+1}\right)$, with energies $\omega_{\pm} = J/2 \pm h_{+}$. The energies of the two-kink states are then obtained as the sum of the energies of the two modes, one from each of the left and right kinks: $\omega_{\pm\pm} = J \pm 2h_{+}$ and $\omega_{\pm}^\text{sym/anti-sym} = J$, where the latter two correspond to symmetric and antisymmetric combinations of $\ket{p}^+_\mathrm{L} \otimes \ket{p}^-_\mathrm{R}$ and $\ket{p}^-_\mathrm{L} \otimes \ket{p}^+_\mathrm{R}$. In the four-kink calculation, $\omega_{\pm}^\text{sym}$ is visible in $I(\omega)$ at $\omega = J$ for the entire range of $h_y$ and $\lambda$ [Fig.~\ref{fig:yz}(g--j)].

Special treatment is necessary when the two kinks are neighboring each other, as the presence of one kink may further restrict the movement of the other. The single flipped spin at site $2p$ (i.e., the two-kink state $\ket{i, l} = \ket{2p,1}$) can only hop to site $2p-1$, that is, to $\ket{2p - 1, 1}$, via the intermediate two-spin-flipped state $\ket{2p-1,2}$, and is disconnected from the rest of the two-kink states. The exact eigenvalues of this three-state system (in contrast to the four states present for decoupled kinks) are $\varepsilon_\pm = J \pm \sqrt{2}h_{+}$ and $\varepsilon_0 = J$, with corresponding wave functions
\begin{equation}
\begin{split}
\ket{\varepsilon_\pm, p} = & \frac{1}{2}\left(\ket{2p-1,1} \pm \sqrt{2}\ket{2p-1,2} + \ket{2p,1}\right), \\
\ket{\varepsilon_0, p} = & \frac{1}{\sqrt{2}}\left(\ket{2p-1,1} - \ket{2p,1}\right).
\end{split}
\end{equation}
As shown in Fig.~\ref{fig:tfim_yz_detail}(a), the $\ket{\varepsilon_\pm}$ states are the only visible states in the linear response $S(\omega)$, as $\varepsilon$ bound states are the only two-kink states that contain a finite single spin-flipped component. $\varepsilon_0$ is invisible at zero momentum transfer, as its wave function is antisymmetric under inversion and thus omitted by the selection rule. Note that though peaks at $\omega = J$ are visible in $I(\omega)$ [Fig.~\ref{fig:tfim_yz_detail}(b)], that should be understood as arising from $\ket{Q}$ being $\omega_{\pm}^\text{sym}$, not $\varepsilon_0$.

Moving beyond the two-kink physics discussed in Ref.~\cite{Woodland2023}, the eigenstates of the four-kink sector can be constructed in a similar manner. When two domains, each defined by a pair of kinks, are far apart, their energies simply add, so the total energy of the four-kink state is the sum of two independent two-kink domains. When the two domains are next to one another, however, their dynamics become coupled, giving rise to new types of bound states. The local Hamiltonian in this sector is spanned by five basis configurations
\begin{equation*}
\begin{split}
\big\{&\ket{2p-1,1,2p+1,1},\, \ket{2p-1,1,2p+1,2},\\
      &\ket{2p-1,1,2p+2,1},\, \ket{2p-1,2,2p+2,1},\\
      &\ket{2p,1,2p+2,1}\big\},
\end{split}
\end{equation*}
with hopping occurring only between adjacent configurations. Diagonalizing this five-state system yields eigenstates $\lambda_i$ with energies $2J \pm \sqrt{3}h_+$, $2J \pm h_+$, and $2J$, where $i = 1,\dots,5$ (ordered by increasing energy). Among them, $\lambda_2$ and $\lambda_4$ are antisymmetric under inversion and are therefore invisible in $I(\omega)$. The lowest and highest states, $\lambda_1$ and $\lambda_5$, are particularly noteworthy: their energies $(2J \pm \sqrt{3}h_+)$ cannot be understood as simple sums of two two-kink bound states. In the literature, such four-kink bound states are sometimes referred to as {\it tetraquark} states~\cite{Vovrosh2022}---the terminology reflecting an analogy to high-energy physics, where mesons (two-quark bound states) can bind into higher composites.

In the experimental THz spectrum of \CoNbO, the characteristic spectral signature of the localized limit, namely the pinch point of the lower-energy branch ($\varepsilon_{-}$) [around 3~T in Fig.~\ref{fig:tfim_yz_detail}(a)], has indeed been observed~\cite{Morris2021}---see also Fig.~\ref{fig:localized_limit}(c) for a reproduction of the experimental spectrum. In the following, we discuss how probing 2DCS in this localized regime can reveal distinct signatures of the tetraquark states. 

\begin{figure}
	\centering
	\includegraphics[width=1.0\columnwidth]{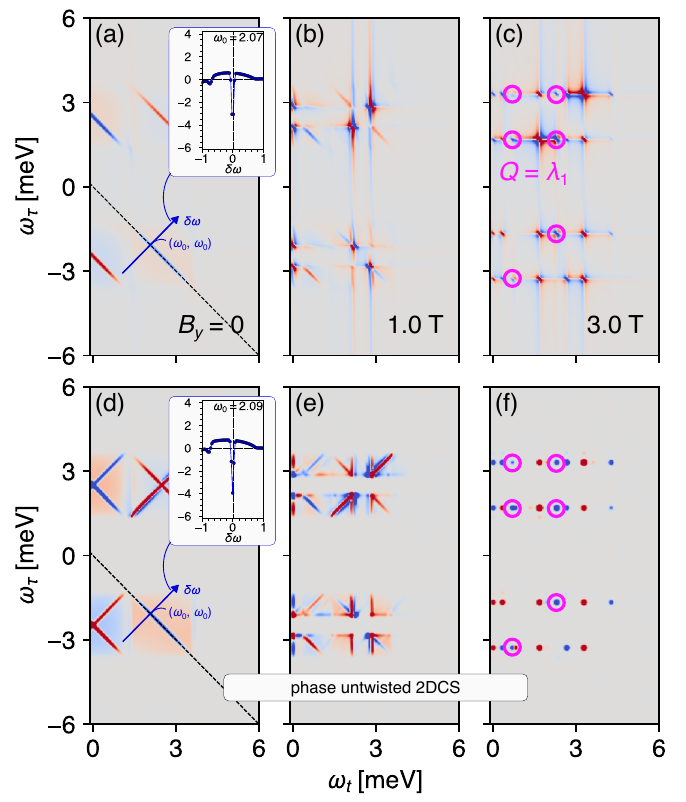}
	\caption{{\bf 2DCS spectra of TFIM + YZ.} (a--c): experimentally accessible signals for each magnetic field. (d--f): corresponding phase-untwisted spectra. A diagonal line cut at $(\omega_t, \omega_\tau) = (\omega_0 + \delta\omega, \omega_0 + \delta\omega)$ is shown in the insets of (a) and (d). Close to the localized limit ($B_y = 3.0$~T), peaks originating from the intermediate state $\ket{Q}$ being the tetraquark $\lambda_1$ (see also Fig.~\ref{fig:tfim_yz_detail}), which also appears in the full \CoNbO\  spectrum, are marked by pink circles.}
	\label{fig:tfim_yz}
\end{figure}

\subsubsection*{2DCS spectra}

Turning to the 2DCS signatures of the TFIM+YZ model ($\lambda = 1.0$), we first consider the exactly solvable point at $h_y = 0$ ($B_y = 0$). The resulting spectrum shows close similarity to that reported for the twisted Kitaev-chain model~\cite{Sim2023_2}: continuous line features (along both the diagonal and anti-diagonal directions) appear at the R/NR locations, as well as at the TR location [Fig.~\ref{fig:tfim_yz}(a, b)]. This similarity can be traced to the fact that both models admit an exact description in terms of non-interacting spinon pairs, with a two sublattice structure, at zero field. This leads to two distinct spinon bands, i.e., $\xi^{\pm}_k$, and hence to a similar structure of the excitation spectrum (with the twisted Kitaev chain having additional complications due to non-monotonic $k$ dependence). The two-spinon continuum spans the energy range $[J(1 - 2\alpha_{yz}), J(1 + 2\alpha_{yz})] \approx [1.35,\,3.60]$~meV.

The anti-diagonal line feature at the R location (i.e., the spinon-echo signal) can be understood in the same way as in the 1$d$-TFIM case: the intermediate states $\ket{P}$, $\ket{Q}$, and $\ket{R}$ correspond either to the ground state or to two-spinon states with momenta $(k,-k)$. Summing over all allowed $k$ values in the two-spinon continuum gives rise to the spinon-echo feature. In the present case, however, despite the absence of interactions, the spinon-echo signal (with negative sign) is accompanied by a broad background square feature (with positive sign) spanning the region $(\omega_t,\omega_\tau)\in[1.35,\,3.60]\times[1.35,\,3.60]$~meV. Similar broad features were reported for the 1$d$-TFIM in Ref.~\cite{Wan2019} when the $\chi^{(3)}_{yyyy}$ response was studied in the presence of a transverse $x$-field (whereas in the previous section we discussed $\chi^{(3)}_{yyyy}$ in a transverse $y$-field). The origin of these broad features can be traced to the fact that, although the model is exactly solvable and hosts non-interacting spinon excitations, the pulse operator $M^y$ does not take a simple form in terms of the Bogoliubov quasiparticles that diagonalize the Hamiltonian. As a result, multiple pathways involving different choices of intermediate states $\ket{P}$, $\ket{Q}$, and $\ket{R}$ contribute to the response, and the signal cannot be understood solely as a sum over pathways involving a single $(k,-k)$ spinon pair. Equivalently, the broad features can be viewed as arising from pathways in which some of $\ket{P}$, $\ket{Q}$, and $\ket{R}$ correspond to different two-spinon states with distinct momenta.

Other diagonal and anti-diagonal sharp line features on top of broad square backgrounds appearing at the TR and NR locations can be understood in a similar manner. The relative sharpness of these lines reflects the fact that the contribution is strongest when all intermediate states $\ket{P}$, $\ket{Q}$, and $\ket{R}$ originate from the same two-spinon state with momenta $(k,-k)$. Note that, in the twisted Kitaev chain~\cite{Sim2023}, the anti-diagonal features at the TR and NR locations exhibit a finite curvature. By contrast, the straight features in the present model follow from the simple relation between the two spinon bands; for example, here $\xi_k-\eta_k = 2\alpha_{yz}\cos(k)$, which has essentially the same $k$-dependence as $2\xi_k$ and $2\eta_k$. In more general cases, such relations need not be so simple, and the anti-diagonal structures can become curved. As we will see later, a similar curved structure appears in the full \CoNbO\ model.

The localized limit $h_y = J\alpha_{yz}$ ($B_y \approx 3$~T), offers a clear perspective on the overall structure of the 2DCS spectrum in a magnetic field. As illustrated in Fig.~\ref{fig:tfim_yz_detail}(d), peaks can be categorized based on the intermediate state $\ket{Q}$, whose locations in the two-dimensional frequency plane follow directly from inserting the energies of the intermediate $\ket{P}$, $\ket{Q}$, and $\ket{R}$ states into Eq.~(\ref{eq:chi3}). Here, $\ket{P}$ and $\ket{R}$ are always either $\varepsilon_{+}$ or $\varepsilon_{-}$ states, which are the only two-kink states visible in $S(\omega)$. The numerically obtained 2DCS spectrum at the localized limit [Fig.~\ref{fig:tfim_yz_detail}(e)] closely matches this schematic picture. In particular, the tetraquark states $\lambda_1$ and $\lambda_5$ become visible in the 2DCS signal. The signature of the $\lambda_1$ mode, marked by pink circles in Figs.~\ref{fig:tfim_yz}(c) and \ref{fig:tfim_yz}(f), is clearly seen in both the experimental and phase-untwisted spectra. As we will see below in the context of \CoNbO, it persists even away from the Ising+YZ limit.

Away from the localized limit [$B_y = 1.0$~T; Figs.~\ref{fig:tfim_yz}(b) and \ref{fig:tfim_yz}(e)], the possible pathways that can give rise to sharp 2DCS signatures become more diverse. As seen in $S(\omega)$ [Fig.~\ref{fig:tfim_yz_detail}(a)], the states $\ket{P}$ and $\ket{R}$ can now also lie within the spinon continuum, leading to additional elongated streak features emanating from the peaks where $\ket{Q}$ corresponds to two $\varepsilon$ modes. Although a four-kink bound state still exists away from the localized limit, as evidenced in $I(\omega)$ [Fig.~\ref{fig:tfim_yz_detail}(b)], the corresponding peak in the 2DCS spectrum now lies in close proximity to these elongated streaks, making it less distinguishable, especially in the phase-twisted spectrum.

\section{Realistic Model for $\textrm{CoNb}_2\textrm{O}_6$}
\label{sec:full_spectrum}

\begin{figure}
	\centering
	\includegraphics[width=1.0\columnwidth]{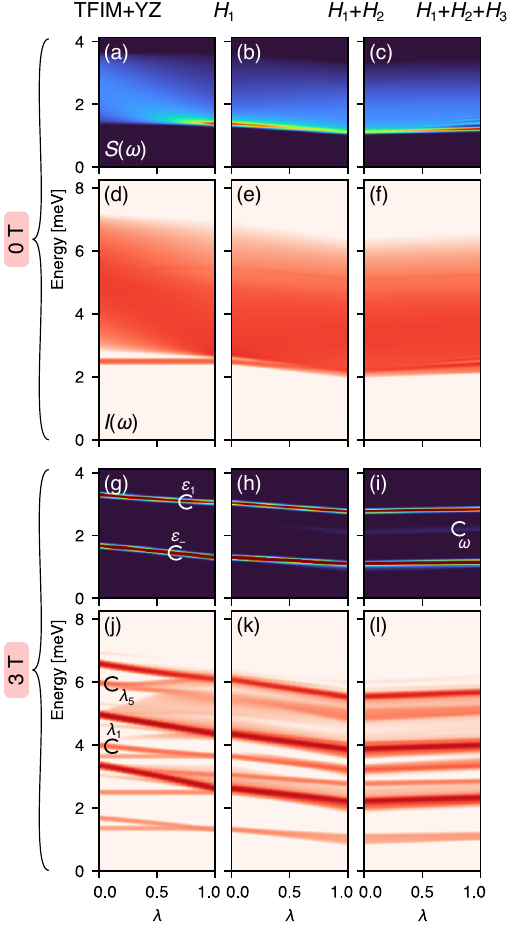}
	\caption{{\bf Spectral evolution towards the \CoNbO\ limit.} Starting from the Ising+YZ model in the leftmost column, we interpolate towards the full \CoNbO\ model in the rightmost column via the sequence TFIM+YZ $\rightarrow \mathcal{H}_1 \rightarrow \mathcal{H}_1 + \mathcal{H}_2 \rightarrow \mathcal{H}_1 + \mathcal{H}_2 + \mathcal{H}_3$, with each segment parameterized by $\lambda \in [0,1]$. (a--c) $S(\omega)$ and (d--f) $I(\omega)$ at $B_y = 0.0$. (g--i) $S(\omega)$ and (j--l) $I(\omega)$ at $B_y = 3.0$~T.}
	\label{fig:H1_to_H3}
\end{figure}

Finally, we present the 2DCS spectra for the full model of \CoNbO. We begin by examining how the linear-response spectra $S(\omega)$ and $I(\omega)$ evolve as we sequentially incorporate the subdominant interactions present in \CoNbO, as shown in Fig.~\ref{fig:H1_to_H3}, in order to build a physical understanding of the full model. We then discuss how these features manifest in the 2DCS spectra.

Starting from the Ising+YZ model, which we developed a good understanding of in the previous section, we observe that the main impact of turning on the XY term to obtain the minimal $\mathcal{H}_1$ Hamiltonian is to shift the overall energies. At $B_y=0.0$~T [Figs.~\ref{fig:H1_to_H3}(a, d)], the continuum of the Ising+YZ model is renormalized, with spectral weight shifted towards lower energies. This trend mirrors the effect of the XY term in the 1$d$-TFIM, where increasing XY strength first transfers weight toward the lower edge of the continuum and, for sufficiently strong XY coupling, can eventually pull the single-SF bound state out of the continuum. In the present parameter regime, we observe the former but not the latter: the XY term produces a substantial renormalization of the continuum, but it does not stabilize a single-SF bound state. Moreover, as shown in Figs.~\ref{fig:H1_to_H3}(g, j), many of the bound states found in the \emph{localized limit} ($B_y \approx 3$~T)---where the dynamics of the kinks are highly restricted due to the interplay between the transverse field and the YZ term---remain localized along this interpolation, although some modes evolve into the continuum, as is most clearly seen in $I(\omega)$ around 5~meV at $B_y = 3.0$~T.

Among the localized excitations, particular attention is paid to the \emph{tetraquark} states, the four-kink bound states $\lambda_1$ and $\lambda_5$ marked in Fig.~\ref{fig:H1_to_H3}(j). Being genuine four-kink bound states, within the kink-conserving approximation these modes are not visible in the conventional linear response $S(\omega)$, which is sensitive only to two-kink states, but they do appear in the higher-order response $I(\omega)$. The $\lambda_1$ mode shifts to lower energy as one approaches $\mathcal{H}_1$ and moves toward the weakly dispersive two-kink state $\omega_{++}$, which can be viewed as a two-kink state where the two kinks are relatively far apart~\footnote{In the four-kink calculation, $\lambda_1$ and $\omega_{++}$ belong to different sectors---four-kink and two-kink, respectively---and therefore do not hybridize.}. Throughout this interpolation, $\lambda_1$ remains clearly visible in $I(\omega)$. In contrast, the $\lambda_5$ mode starts to overlap with the $2\varepsilon_{+}$ branch, which may be interpreted as two independent two-kink bound states (i.e., four kinks in total). This near-degeneracy may lead to strong hybridization, making it difficult to unambiguously identify this mode.

The additional contributions from $\mathcal{H}_2$ and $\mathcal{H}_3$ have only a minor effect on the linear-response spectra. Near the localized limit, their main role is to slightly shift the overall spectrum [Fig.~\ref{fig:H1_to_H3}]. At $B_y = 0.0$~T, the effective longitudinal field introduced by $\mathcal{H}_3$ leads to a discretization of the continuum signal due to confinement-induced bound state formation. In the 1$d$-TFIM, the longitudinal field is known to produce two effects in $S(\omega)$: (1) it shifts spectral weight toward the lower edge of the continuum, and (2) it discretizes the spectrum. In our case, the first effect is already present in the full $\mathcal{H}_1$ model, so the main contribution of $\mathcal{H}_3$ is to induce a mild discretization of the spectrum.

\subsection{High-temperature model ($\mathcal{H}_1$+$\mathcal{H}_2$)}

\begin{figure}
	\centering
	\includegraphics[width=1.0\columnwidth]{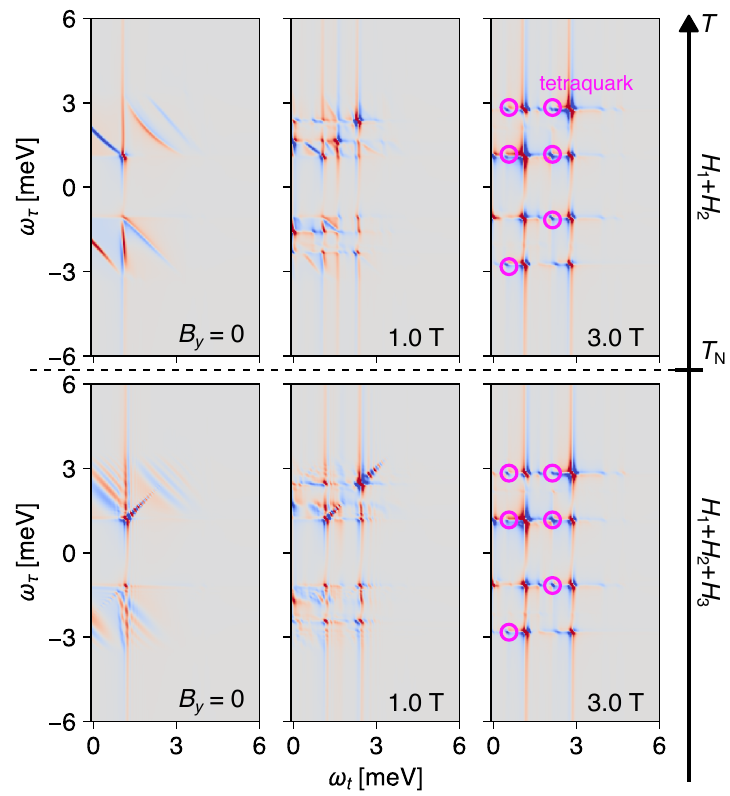}
	\caption{{\bf 2DCS spectrum of \CoNbO} above and below $T_\mathrm{N}$. The upper panels show the spectra for the high-temperature regime, modeled by $\mathcal{H}_1 + \mathcal{H}_2$, while the lower panels correspond to the low-temperature regime with the additional spinon-confining potential $\mathcal{H}_3$. From left to right, $B_y = 0.0,\, 1.0,\, \text{and}\, 3.0~\text{T}$, respectively.}
	\label{fig:H1H2_H1H2H3_2DCS}
\end{figure}

Now we move on to the 2DCS spectroscopic signatures of the high-temperature phase above $T_\mathrm{N}$ [Fig.~\ref{fig:H1H2_H1H2H3_2DCS}], which we mimic by ignoring the interchain coupling term $\mathcal{H}_3$. This treatment has been shown~\cite{Morris2021} to capture the essential features of the linear-response spectra of \CoNbO\ above $T_\mathrm{N}$.

At $B_y = 0.0$~T, the spectrum closely resembles that of the exactly solvable Ising+YZ model: line-like features appear at the NR and TR signal positions, together with an anti-diagonal \emph{spinon-echo} feature at the R position, which constitutes a characteristic 2DCS signature of fractionalized excitations in the two-dimensional frequency plane. However, there are several notable differences. First, the spinon-echo signal has enhanced spectral weight near the lower edge of the spinon continuum. 
We also see another feature already observed in the Ising+XY model, namely, vertical streaks emerging from the lower edge of the continuum in the R signal. A further deviation from the Ising+YZ case is that the NR and TR signals exhibit anti-diagonal features with finite curvature. This behavior resembles that of the twisted Kitaev chain model~\cite{Sim2023_2, Watanabe2024}, where the curvature arises from having two distinct spinon bands ($\xi_k$ and $\eta_k$)---originating from a unit cell with two sites---with non-monotonic $k$ dependence of the band splitting $\xi_k-\eta_k$.
 
At $B_y = 3.0$~T, despite the absence of an explicit confining potential, the overall structure of the 2DCS spectrum is already relatively discrete due to localization effects inherited from the Ising+YZ model. Most importantly, the signature of the tetraquark state $\lambda_1$ seen in $I(\omega)$ is clearly visible here as well and is marked by pink circles in the figure.

\subsection{Low-temperature model ($\mathcal{H}_1$ + $\mathcal{H}_2$ + $\mathcal{H}_3$)}

In modeling the low-temperature spectroscopic signatures of \CoNbO\  we need to consider the full model introduced in Sec.~\ref{sec:model}, i.e.\ now also include the interchain coupling which we had omitted to discuss the high-temperature signatures above. The 2DCS spectra for the full model of \CoNbO\  are shown in Fig.~\ref{fig:H1H2_H1H2H3_2DCS}. The main impact of the interchain coupling term $\mathcal{H}_3$ is to discretize the continuum spectrum due to spinon confinement. However, this effect is barely noticeable near the localized limit at $B_y = 3.0$~T. As a result, the 2DCS spectra appear almost identical to those without $\mathcal{H}_3$, except for a slight overall energy renormalization. The signature of the tetraquark state $\lambda_1$ remains clearly visible, as marked by circles in the figure.

At $B_y = 0.0$ T, the impact of $\mathcal{H}_3$ on the linear-response spectra $I(\omega)$ is still minor, apart from the expected discretization. This is due in part to the high density of four-kink states in this regime. In contrast, $\mathcal{H}_3$ has a more pronounced effect on the 2DCS spectra. First, due to the discretization, certain diagonal features at the NR signal position become more prominent. These features were previously obscured by phase-twisting effects in the continuum signal limit. Second, the relatively sharp anti-diagonal structure observed in the R signal is no longer present. Instead, it is replaced by a broad, sign-alternating oscillation. 

This is similar to the case of the 1$d$-TFIM in a weak longitudinal field, where the sharp R signal evolves into weak oscillations in the small-field limit~\cite{Watanabe2024}. In contrast, increasing the longitudinal field leads to a grid-like structure formed by additional off-diagonal peaks. In that case, the grid pattern is accompanied by finer features that were interpreted as signatures of domain-wall splitting, in a regime where the confined spinon picture remains valid. The vertical streak originating from the lower edge of the continuum in the R signal remains visible.

\begin{figure}
	\centering
	\includegraphics[width=1.0\columnwidth]{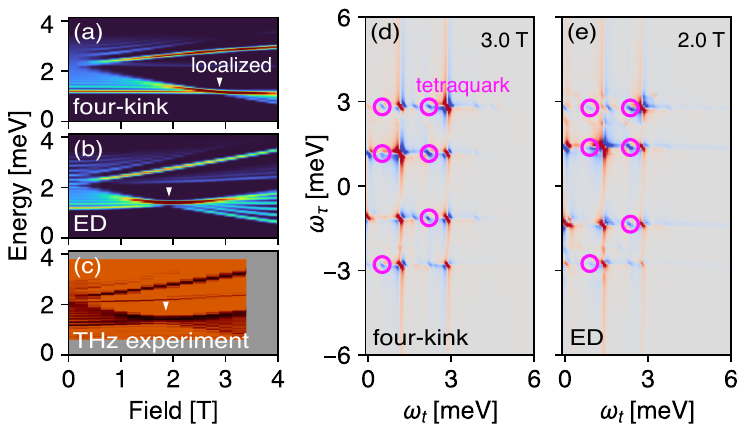}
	\caption{{\bf Experimental identification of the tetraquark signature in the localized limit.} 
	(a--c) Linear-response spectra $S(\omega)$ obtained from four-kink calculations (a), ED (b), and experimental THz spectroscopy data measured at $T = 1.5~\text{K} < T_\text{N}$ and taken from Ref.~\cite{Morris2021} (c). For each plot, the localized limit---defined as the field at which the lower branch of the two-spinon continuum exhibits a pinch point (around $B_y = 3.0$~T in the four-kink calculation and $2.0$~T in ED and experiment)---is indicated by an inverted triangle. (d, e) 2DCS spectra obtained from the four-kink calculations and ED, respectively, at the corresponding localized limits, where the tetraquark signatures are marked by pink circles.}
	\label{fig:localized_limit}
\end{figure}

Let us conclude this section by discussing the experimental observability of the tetraquark signatures in \CoNbO. So far, our analysis has relied on the four-kink approximation which, while offering good numerical scalability and a transparent theoretical picture, is approximate by construction. It is therefore natural to ask whether the tetraquark signatures remain robust once we go beyond this approximation and can thus be expected to appear in experiment. To address this question, we compare the four-kink results with ED, which captures effects beyond the four-kink truncation. As shown in Fig.~\ref{fig:localized_limit}(a--c), the localized limit can be identified in the four-kink calculation, in ED, and in the experimental THz spectroscopy data by the presence of a pinch point in the lower branch of the two-spinon bound states in $S(\omega)$. The agreement between the ED and experimental data is excellent, confirming that the localized limit in \CoNbO\ occurs around $B_y \approx 2.0$~T. Although the precise field at which this occurs differs slightly between ED and the four-kink calculation, an appropriate rescaling of the magnetic field allows us to compare the corresponding 2DCS spectra at their respective localized limits [Figs.~\ref{fig:localized_limit}(d, e); see also Appendix~\ref{app:ED_vs_4kink}]. In both calculations, the overall 2DCS structure is very similar, and the tetraquark signature remains clearly visible, as indicated by the pink circles. This suggests that analogous signatures should be observable in future 2DCS experiments on \CoNbO\ below $T_\mathrm{N}$.

\section{Summary and Outlook}
\label{sec:summary_outlook}

\subsubsection*{2DCS spectra}

We have presented 2DCS spectra for a realistic model of \CoNbO, as well as for a variety of related one-dimensional ferromagnetic models. By interpolating between these models, we traced how continuum features and bound states evolve, shedding light on the origin of key spectral features in \CoNbO.

In the absence of the applied transverse field, $B_y = 0.0$~T, we showed that, for \CoNbO, the anti-diagonal ``spinon-echo" R signal is fragile under the inclusion of the effective interchain coupling term $\mathcal{H}_3$, which acts as a longitudinal field. This situation is analogous to the 1$d$-TFIM where, once the spinon-confining potential of a longitudinal field is introduced, the sharp anti-diagonal spinon-echo feature at the R-signal position is quickly washed out and replaced by weak, sign-alternating oscillations~\cite{Watanabe2024}. The potential observation of a spinon-echo signal in \CoNbO\ is thus likely only above $T_N$, where $\mathcal{H}_3$ can be neglected but finite temperature effects will need to be considered.

Near the localized limit, which, within the four-spinon approximation, occurs around $B_y = 3.0$~T, the overall structure of the 2DCS spectra can be well understood by comparison with the localized-limit spectra of the purely Ising+YZ model (in which the localization is exact within our approximation). We demonstrated that 2DCS provides access to the dynamics of the four-spinon sector, which is not easily accessible through linear-response measurements. In particular, we identified clear signatures of tetraquark states in the 2DCS spectrum, highlighting the advantage of nonlinear spectroscopy in probing interaction effects between excitations and revealing the existence of bound states. Away from the localized limit, the spectrum becomes more complex due to the presence of multiple overlapping continua.

\subsubsection*{Experiments and theoretical refinements}

In this work, we specifically focused on the third-order response function $\chi^{(3;2)}$, but other components such as $\chi^{(3;1)}$ and $\chi^{(2)}$ will also contribute, offering further opportunities for potentially accessing unique physics but also potentially complicating the interpretation of the experimental signatures discussed here. Fortunately, as long as the peak positions of different contributions do not significantly overlap, it should be possible to distinguish their origins by analyzing the THz pulse-intensity dependence of the signal~\cite{Lu2017}. Furthermore, the fourth quadrant of the 2DCS spectrum, where the ``spinon echo'' signals and some of the tetraquark signatures are expected to appear, is relatively free from contamination by other contributions, making it a promising region for experimental exploration.

A more quantitative comparison with future experiments could benefit from improved modeling of how the THz pulses couple to the magnetic degrees of freedom and how the resulting dynamics is mapped onto the measured observable (i.e., the effective coupling/measurement operator; see, e.g., Refs.~\cite{Krupunov2023, Brenig2024, Brenig2025, Brenig2025_2,Srivastava2025}). Here we assumed a dipolar coupling of the THz field to the magnetization, so both pulses enter through the same operator $M$ and the detected signal is also expressed in terms of $M$. Depending on the microscopic coupling and detection scheme, however, the measured observable may also be sensitive to composite channels. In that case, multi-pulse driving can populate the higher-order states discussed here, and their subsequent relaxation may contribute to the emitted signal even at lower order---for example, signatures that we discuss in terms of $\chi^{(3;2)}$ could already appear in $\chi^{(2)}$ if such composite channels are accessible.

\subsubsection*{Broader context}

In the broader context of nonlinear response studies of quantum magnets, an emerging pattern is that nonlinear probes can reveal information that is difficult to access within linear response. This is perhaps unsurprising: the experimental protocol underlying nonlinear spectroscopy involves composite perturbations, i.e., multiple pulses, and therefore naturally couples to processes that are invisible or strongly suppressed at linear order. Examples include probing interaction effects between elementary excitations~\cite{Wan2019, Fava2021, Fava2023}, detecting composite excitations~\cite{Nandkishore2021, Watanabe2025} (including the tetraquark states studied here), and potentially accessing braiding statistics of anyonic excitations~\cite{McGinley2024, McGinley2024_2, Yang2025, Kirchner2025}. In all of these situations, one may envisage disentangling the feature of interest from other contributions by using different pulses---or, equivalently, by working in multi-dimensional time/frequency space---to address different elementary (or anyonic) species and then examining their interaction, fusion (into composite bound states), or braiding response.

Regarding interaction effects among fractionalized excitations and the associated acquisition of a finite lifetime, it has been suggested that one could infer interactions from the broadening of the ``spinon-echo'' signal~\cite{Wan2019, Hart2023}. However, as we found in Secs.~\ref{sec:ising_xy} and \ref{sec:ising_yz}, interaction effects can manifest more sensitively even when the lifetime is (theoretically) infinite, and in some cases even in non-interacting settings, through the detailed structure of the 2DCS spectra. A systematic understanding of how to best exploit this sensitivity remains an interesting direction for future work.

Finally, on the experimental front, while 2DCS measurements are accumulating for magnetic materials with conventional magnon excitations~\cite{Lu2017, Huang2024, Zhang2024, Zhang2024-2, Zhang2025}, applying these techniques to quantum magnets---where fractionalization and multi-particle bound states play a central role---remains an exciting opportunity. Our results provide concrete targets for such experiments by identifying robust, experimentally accessible signatures of composite fractionalized excitations in a realistic model of \CoNbO.

\vspace{5mm}

{\it Data availability.---}
The numerical data shown in the figures are available on Zenodo~\cite{zenodo_repository}.

\acknowledgments
We acknowledge partial funding from the DFG within Project-ID 277146847, SFB 1238 (projects C02, C03). This work was performed, in part, in summer 2024 during the 
``Probing Collective Excitations in Quantum Matter by Transport and Spectroscopy" program
at the Aspen Center for Physics, which is supported by National Science Foundation grant PHY-2210452. The numerical simulations were performed on the JUWELS cluster at the Forschungszentrum Juelich, the Noctua2 cluster at PC$^2$ in Paderborn, and the RAMSES cluster at RRZK Cologne.

\appendix

\section{Four-kink vs ED}
\label{app:ED_vs_4kink}

\begin{figure}[h!]
	\centering
	\includegraphics[width=1.0\columnwidth]{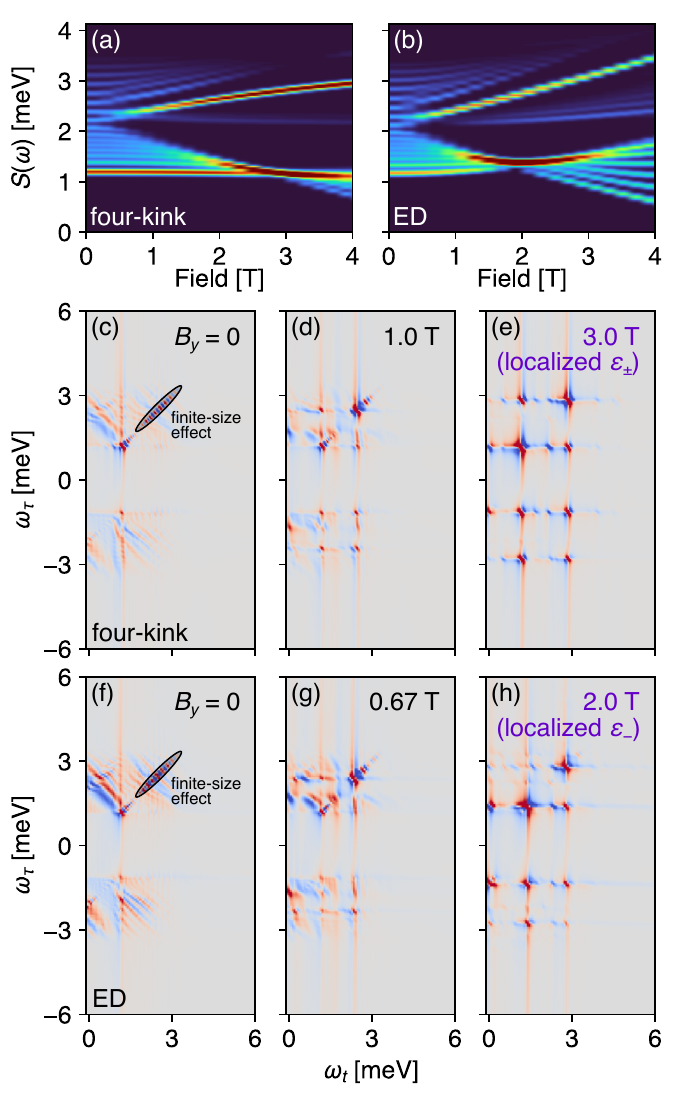}
	\caption{{\bf Comparison of four-kink and ED spectra of \CoNbO\ below $T_\mathrm{N}$} for a system size $L = 20$. (a, b) Linear-response spectra $S(\omega)$ from four-kink and ED calculations, respectively. (c--e) Four-kink 2DCS spectra for $B_y = 0.0,\, 1.0,\, \text{and}\, 3.0~\text{T}$, respectively. (f--h) Corresponding ED spectra. In the ED calculation, we define the localized limit at $B_y \approx 2.0$~T (as opposed to $3.0$~T in the four-kink calculation) as the point where the linear-response spectrum $S(\omega)$ exhibits a pinch-point feature in the $\varepsilon_{-}$ branch.}
	\label{fig:ED_full_model}
\end{figure}

In the main text, the 2DCS spectrum of \CoNbO\ is presented based on the four-kink calculation with system size $L = 100$. In Ref.~\cite{Woodland2023}, its two-kink variant was shown to reproduce the experimental linear-response spectrum, so we expect that the four-kink calculation also captures the essential features of the 2DCS spectrum. Here, we justify this expectation by comparing the four-kink result for $L = 20$ with the ED result for the same system size, for the full model of \CoNbO, as shown in Fig.~\ref{fig:ED_full_model}. Note that the ED calculation reproduces the experimental spectra~\cite{Woodland2023}, including the quantitative field dependence, quite well.

In the linear-response spectrum $S(\omega)$, both methods yield qualitatively similar results, with particularly good agreement near zero field [Figs.~\ref{fig:ED_full_model}(a) and \ref{fig:ED_full_model}(b)]. As the transverse field increases, discrepancies become more pronounced, which likely arises because, in the four-kink calculation, the two-spinon and four-spinon sectors are treated independently, neglecting their coupling. As a consequence, the field at which the pinch-point feature---the characteristic signature of the localized limit---appears differs between the two methods: in the four-kink calculation it appears at $B_y \approx 3.0$~T, with both the upper and lower branches ($\varepsilon_{+}$ and $\varepsilon_{-}$) showing pinch-point features at similar fields, whereas in the ED calculation the pinch point in the lower branch appears at $B_y \approx 2.0$~T, consistent with the THz experiment~\cite{Morris2021}.

Aside from these quantitative differences, the overall structure of the linear-response spectrum is well captured by the four-kink calculation. We also compare the 2DCS spectra obtained from both methods at three different field strengths: $B_y = 0.0,\, 1.0,\, \text{and}\, 3.0$~T for the four-kink calculation, and the corresponding fields scaled by a factor of $2/3$ in ED so that the localized limit matches in both approaches. Both methods show good agreement in the overall structure of the 2DCS spectra, including the presence of the tetraquark signature near the localized limit [Figs.~\ref{fig:ED_full_model}(e) and \ref{fig:ED_full_model}(h)].

\newpage
 
\bibliography{ref}

\end{document}